\documentclass[fleqn,usenatbib]{mnras}

\usepackage{newtxtext,newtxmath}
\usepackage{booktabs}

\usepackage[T1]{fontenc}

\DeclareRobustCommand{\VAN}[3]{#2}
\let\VANthebibliography\thebibliography
\def\thebibliography{\DeclareRobustCommand{\VAN}[3]{##3}\VANthebibliography}

\usepackage{graphicx}	
\usepackage{amsmath}	

\title[Slow jets from obscured accretion]{The link between obscured accretion and mildly relativistic precessing jets}

\author[R.P. Fender \& S.E. Motta]{
Rob Fender,$^{1}$\thanks{E-mail: rob.fender@physics.ox.ac.uk}
Sara Motta,$^{2}$
\\
$^{1}$Astrophysics, Department of Physics, University of Oxford, Keble Road, Oxford, OX1 3RH, UK\\
$^{2}$Istituto Nazionale di Astrofisica (INAF), Osservatorio Astronomico di Brera, via E. Bianchi 46, 23807 Merate (LC), Italy\\
}

\date{Accepted XXX. Received YYY; in original form ZZZ}

\pubyear{\the\year{}}

\begin{document}
\label{firstpage}
\pagerange{\pageref{firstpage}--\pageref{lastpage}}
\maketitle

\begin{abstract}
We have recently shown evidence that the most relativistic jets (with Lorentz factor $>2$) from stellar-mass black holes in X-ray binary systems may be locked to a fixed axis, likely the spin axis of the black hole. Slower, mildly relativistic jets (with velocities typically $\sim 0.3c$) are often seen to precess and can be associated with both neutron stars and black holes. In this paper we demonstrate an additional clear link between highly obscured systems and these lower-velocity, precessing jets. We speculate that this link may be due to mass-loading of the jets close to their launch sites, since these obscured systems are likely to be examples of (sometimes persistent, other times transient) super-Eddington accretion. The fastest relativistic jets are now seen to be both locked to a fixed direction, likely the black hole spin axis, {\em and} to be launched in low-density environments, while jets launched in dense environments are generally slower and very likely to precess.
\end{abstract}

\begin{keywords}
accretion, accretion discs -- ISM:jets and outflows -- X-rays:binaries
\end{keywords}



\section{Introduction}

Relativistic jets from black holes are among the most spectacular signposts of accretion. They are present across the black hole mass range, from stellar-mass black holes in X-ray binary systems (XRBs, \citealt{Fender2016}) to supermassive black holes in Tidal Disruption Events (TDEs, \citealt{Alexander2020}) and Active Galactic Nuclei (AGN, \citealt{Blandford2019}). 

In \cite{FM25} (hereafter FM25) we compiled a sample of directly-measured jet speeds from black hole and neutron star XRBs. In that analysis we found that jets from black hole XRBs (BHXRBs) can show speeds that can range from being mildly relativistic ($\beta \Gamma \sim 0.3$ where $\beta = v/c$ is the jet speed expressed as a fraction of the speed of light and the Lorentz factor $\Gamma = [1-\beta^2]^{-1/2}$) to highly relativistic ($\beta \Gamma >5$). Following FM25, we make a binary distinction between 'fast' (F in Table 1) jets as those with $\beta \Gamma > $ and 'slow' (S) jets as those below this threshold. More recently, Lilje, Fender \& Matthews (2026) have shown that the distribution of speeds of BHXRB jets from that sample is consistent with that from AGN, including the fastest blazars (see also \citealt{Zhang2025}) once the orientation of the targets is properly accounted for. 

In FM25 we argued that the fastest jets were locked to a fixed axis, which was likely to be the spin axis of the black hole. We reached this conclusion as every example of precessing or wobbling jets was associated with mildly relativistic jet speeds. Strong support for this scenario was recently reported in \cite{Jiang2026} who reported that in the archetypal relativistic jet BHXRB, GRS 1915+105, the jets had switched to a new orientation after more than three decades seemingly at a fixed axis, but were also considerably slower ($\beta \Gamma \sim 0.3$) than the historical jets ($\beta \Gamma >2$).

\section{Obscured X-ray binaries}

In compiling the data presented in FM25 and in considering the new results on GRS 1915+105 \citep{Jiang2026}, a possible link between obscured accretion and slow(er) astrophysical jets has become apparent. 
In this work, with the term {\em obscured} we will collectively refer to those sources affected by uniform or patchy local absorption with column density N$_{\rm H} \geq 10^{22}$ cm$^{-2}$, although some of the systems in our sample (e.g. SS433) are formally Compton-thick sources\footnote{Compton-thick sources are affected by column density N$_{\rm H} \geq 1.5 \times 10^{24}$ cm$^{-2}$, corresponding to a Thomson optical depth $\tau \geq 1$. 
}.

For a typical low-mass XRB at a distance of a few kpc and Galactic latitude $|b| \lesssim 5^\circ$, the total Galactic hydrogen column density along the line of sight is $N_\mathrm{H} \sim 10^{21}$--$10^{22}$\,cm$^{-2}$ \citep{HI4PI2016,PlanckXI2014}. This implies that in general, in any given unobscured XRB showing signatures of absorption within the above density range, the contribution of a local absorber must be negligible against the Galactic foreground. The obscured systems in our sample, by contrast, require $N_\mathrm{H}$ values that are clearly in excess of the Galactic column (in several cases by more than an order of magnitude) strongly suggesting the presence of a dense, possibly variable local absorber, associated with processes in the systems itself (winds/outflows from the accretion disc and/or the donor star). We stress that 
the systems discussed in the next section are the only ones we are aware of with evidence for variable local absorption.
In the following we discuss each of such obscured XRBs individually. 

\subsection{Obscured X-ray binaries with measured jet speeds}

There are five XRBs with evidence for strong local obscuration which also have measured jet speeds. We discuss these below.

\subsubsection{SS433}
The archetypal XRB with slow, precessing jets is SS433, which has jets precessing at $\beta \Gamma \sim 0.3$ with a 162.5 day period \citep{Margon1984a}.  This system is heavily obscured by its own inflated accretion disc and associated wind (a consequence of sustained super-Eddington accretion). The observed X-ray luminosity is believed to be less than 1\% of the intrinsic luminosity \citep{Khabibullin2016}. It is widely thought that SS433 would be considered an ultra luminous X-ray source (ULX) if viewed closer to the jet/accretion funnel axis \citep{Begelman2006, Middleton2021} and it has been suggested that it is a small scale local analogue to the obscured accretion taking place in 'Little Red Dots' (LRDs) at high redshift \citep{ZhangL2025}.

\subsubsection{V404 Cyg}
The 2015 outburst of the relatively nearby BH XRB V404 Cyg generated much excitement, with an expectation that we would get exquisite views of highly relativistic jets. It had already been noted in its 1989 outburst that there was evidence for highly variable X-ray absorption local to the source \citep{Zycki1999}, and this was spectacularly confirmed in the 2015 outburst \citep[e.g.][]{Motta2017a, Motta2017b}. However, the jets observed from V404 Cyg were strongly precessing and only mildly relativistic \cite{Miller-Jones2019} and despite intensive observations covering a large number of radio flares \citep{Fender2023} no highly relativistic jet was ever found.

\subsubsection{Cyg X-3}

Cyg X-3 is another highly obscured XRB, which stands apart from other obscured X-ray binaries because the heavy obscuration is largely (but not uniquely) due to strong stellar winds from the Wolf-Rayet companion star. The stellar wind extends far beyond the (short, 4.8 hr) binary orbit, thus Cyg X-3 is constantly embedded in a high-density environment \citep{Szostek2008}. Recently, IXPE polarimetric data have shown that Cyg X-3's central engine is completely hidden behind an optically thick envelope shaped as a narrow funnel, most likely sustained by super-Eddington accretion \citep{Veledina2024, Mikusincova2025}. 
The strong local absorption is beautifully illustrated by the orbital modulation of the X-ray emission at the 4.8 hr binary period; see e.g. \citealt{Mason1979}. The system has shown both slow and fast jets, as well as evidence for precession \citep{Geldzahler1983, Schalinski1995, Mioduszewski2001}. Cyg X-3 may be an example of a source where the fast jet sometimes manages to 'break out' of the local obscuration and sometimes are stifled by it.

\begin{table*}
\centering
\caption{Selected jet properties for the sample of sources in Fender \& Motta (2025; to which the reader should refer for the whole table), adding a column indicating if accretion in the system is significantly obscured by local material. Also different to FM25 are (i) GRS 1915+105 now has two entries, one for its pre-obscured state before 2019 and the other for its obscured state since (the former, fast/unobscured state corresponds to the entry in FM25), (ii) Hughes et al. (2026 in prep) have demonstrated that deceleration model fits do {\em not} constrain the Lorentz factor for MAXI J1348-630. 
Speed classification: F = fast ($\beta\Gamma \geq 1.0$); S = slow ($\beta\Gamma < 1.0$);
$>$F = fast with $\beta_\mathrm{intrinsic}$ capped at 0.9 ($\Gamma \geq 2.3$, $\beta\Gamma \geq 2.0$).
Direction: L = locked; P = precessing; ? = undetermined.
Obscured: Y = yes; N = no; ? = unknown.
Note that in cases where a classification letter is followed by "?" (e.g. "L?") then in our quantitative analysis we use that classification indicated by the letter and the question mark is there to indicate some small uncertainty in the classification.
}
\label{tab:jets}
\begin{tabular}{lccccc}
\toprule
Name & $\beta\Gamma$ & Accretor & Speed & Direction & Obscured? \\
\midrule
4U\,1543$-$47            & $>$5.2  & BH  & $>$F & L   & N \\
GX\,339$-$4              & $>$2.06 & BH  & $>$F & L   & N \\
XTE\,J1550$-$564         & $>$2.06 & BH  & $>$F & ?   & N \\
MAXI\,J1803$-$298        & $>$2.06 & BH  & $>$F & ?   & N \\
MAXI\,J1820$+$070 (fast) & $>$2.06 & BH  & $>$F & ?   & N \\
MAXI\,J1820$+$070 (slow) & 0.31    & BH  & S    & ?   & N \\
MAXI\,J1348$-$630        & $>$2.06  & BH  & $>$F     & L   & N \\
GRS\,1915+105 (pre-2019)        & $>$2.00 & BH  & $>$F & L   & N \\
GRS\,1915+105 (post-2019)        & $0.37$ & BH  & S & P   & Y \\
GRO\,J1655$-$40          & $>$2.06 & BH  & $>$F & L?  & N \\
MAXI\,J1535$-$571        & 1.09    & BH  & F    & ?   & N \\
H1743$-$322              & 1.36    & BH  & F    & ?   & N \\
MAXI\,J1848$-$015        & 1.08    & BH? & F    & ?   & N \\
XTE\,J1752$-$223         & 0.77    & BH  & S    & L?  & N \\
EXO\,1846$-$031          & 0.39    & BH  & S    & ?   & N \\
V404\,Cyg                & 0.48    & BH  & S    & P   & Y \\
XTE\,J1908+094           & 0.14    & BH  & S    & ?   & N \\
SS433                    & 0.25    & ?   & S    & P   & Y \\
Cyg\,X-3 (fast)          & 1.68    & ?   & F    & L?  & Y? \\
Cyg\,X-3 (slow)          & 0.59    & ?   & S    & P   & Y \\
Cir\,X-1                 & 1.44    & NS  & S    & P   & Y? \\
Sco\,X-1                 & 0.62    & NS  & S    & P   & N? \\
Cyg\,X-2                 & 0.32    & NS  & S    & ?   & N \\
\bottomrule
\end{tabular}\label{tab:data}
\end{table*}

\subsubsection{Cir X-1}

Cir X-1 is the youngest known X-ray binary, sitting within its natal supernova remnant, and is the only one of our obscured sample which definitely contains a neutron star (although the natures of the accretors in SS433 and Cyg X-3 remain unclear, due to the obscuration). Similarly to SS433, the slow precessing jets from the binary are shaping the remnant blowing into it bubbles aligned with the jet direction (Gasealahwe, Savard et al. 2025; Cowie et al. 2026). 
In Cir X-1 strong disc winds driven by erratic super-Eddington accretion cause variable obscuration of the system \cite{Clark2003} likely at periastron passage in a 16.6-day highly eccentric orbit. Winds from the companion star may also contribute to variable absorption \citep{Yu2024}. 

\subsubsection{GRS 1915+105}
GRS 1915+105 was the first XRB in our galaxy to reveal apparent superluminal motion \citep{Mirabel1994}. This system is the archetypal BH XRB for relativistic jets, and indeed for the connection between accretion states and modes of jet production \citep{Fender2004}. Historically the system has shown multiple ($>10$) fast ejections at a fixed axis since 1994, and no evidence for strong local absorption. However, in 2019 the system entered a state of reduced X-ray luminosity accompanied by strong radio flaring, indicating jet activity in a highly obscured phase \citep{Motta2021, Miller2020, Sanchez-Sierras2023, Miller2025}, which persists to this day. Remarkably, \cite{Jiang2026} have shown the jets in this phase have changed dramatically in direction at the same time as dropping in speed, in strong confirmation of the hypothesis proposed in FM25. This appears to be a case of a jet system switching from unobscured accretion producing fast jets at a fixed axis, to highly obscured accretion producing slower, precessing, jets.

\subsection{Obscured X-ray binaries not included in the analysis}

There are four XRBs with evidence for strong variable local absorption that do not have measured jet speeds in the same period of time, some of which however may be good candidates for future measurements. We discuss these below.

\subsubsection{V4641 Sgr}
V4641 is an unusual, obscured, XRB which stands out for its likely super-Eddington accretion, variable obscuration, and radio flaring which are believed to be associated with an ultra-high-energy gamma-ray bubble surrounding the source \citep{Alfaro2024}. Similarly to V404 Cyg, super-Eddington X-ray emission likely drives a strong disk winds, which expel large amounts of material that surrounds the system \citep[][]{Koljonen2020}. 

A recent reanalysis of the radio imaging taken during the 1999 outburst of the system by \cite{Marti2026} reveals a clear two-sided radio jet. However, speed estimates are hampered by few epochs and ambiguous launch date and we are not at present able to add it to our sample in a quantitative way (hence it is not included in Table 1 nor subsequent analyses).

\subsubsection{GX 13+1}

GX 13+1 is a NS XRB classified as a Z source, persistently accreting near the Eddington limit, similar to Sco X-1 and, at times, Cir X-1. Recent XRISM Resolve observations caught the source during a super-Eddington episode, revealing one of the densest winds ever seen in X-ray absorption — a Compton-thick outflow that significantly attenuates the observed flux, making the source appear intrinsically faint despite accreting at or above the Eddington limit \citep{Xrism2025}. GX 13+1 is radio-bright and variable e.g. \citep{Homan2004}, but to date has no resolved jets.

\subsubsection{GRO J1655-40 in 2005}
 GRO J1655-40 is a well-studied BH XRB which has undergone two major outbursts (in 1994 and 2005). During the 2005 outburst, Chandra HETGS observations revealed a dense, highly ionized accretion disc wind \citep{Miller2006c, Miller2008}. \cite{Neilsen2016} argued that the unusual properties of this phase were consistent with a super-Eddington, Compton-thick wind obscuring the inner accretion flow. However, the resolved jet speed measurements for this system are only associated with the 1994 outburst \citep{Hjellming1995}, which does not show evidence for such obscuration. 

\subsubsection{Her X-1}
Finally, we note that the  Her X-1, arguably one of the most studied precessing XRB jet system (with a 35-day super-orbital period driven by radiation-pressure-induced disc precession; e.g. \cite{Ogilvie2001}), is not included in our sample as no resolved jet speed measurement exists for this system. Indeed, although it has been detected as a weak radio source, the origin of this radio emission remains unclear \citep{vandenEijnden2018}.

\begin{figure*}
	\includegraphics[width=2\columnwidth]{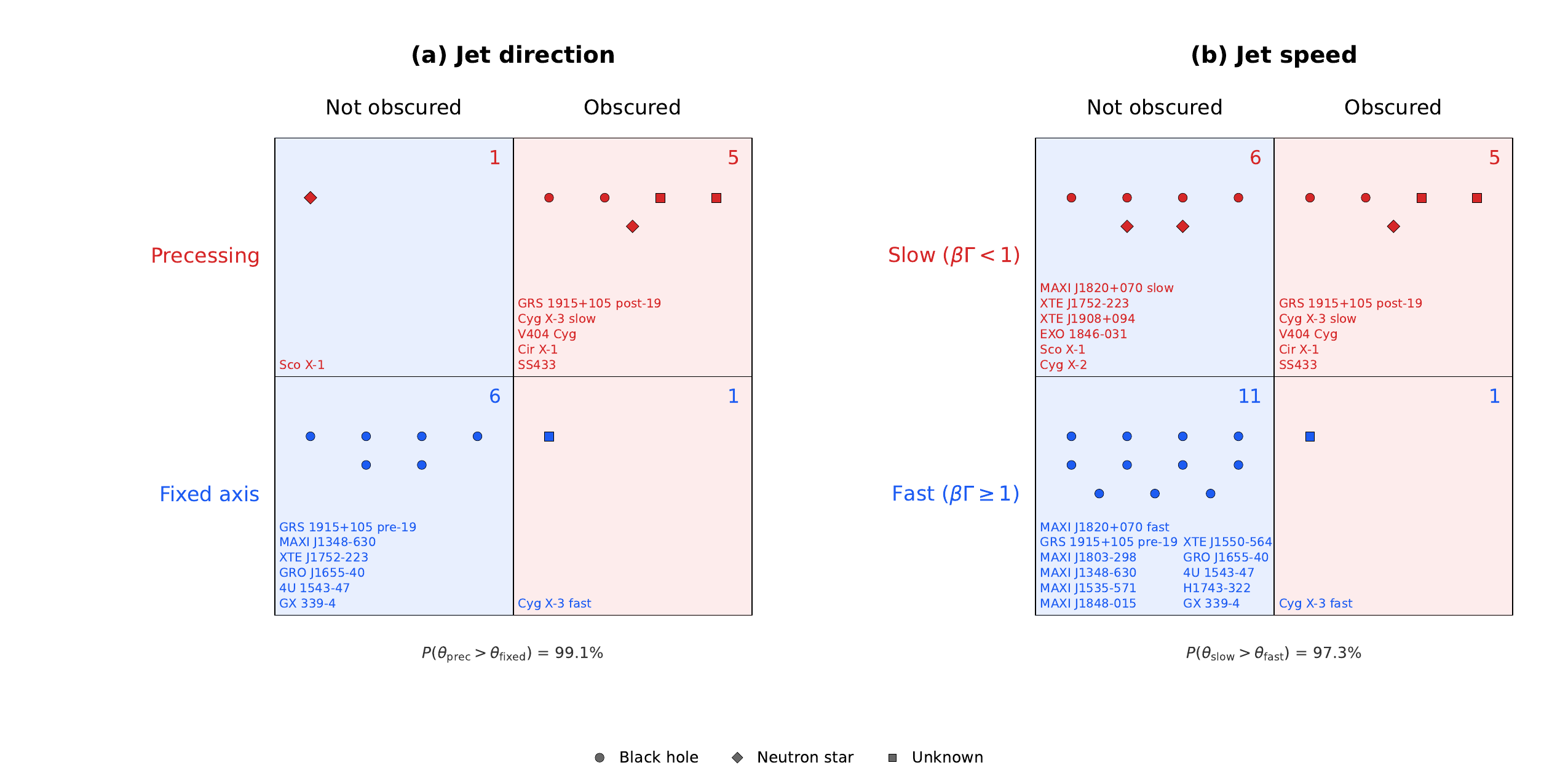}
  \caption{
  Our classifications and sources for testing the correspondence between obscured accretion and precessing/fixed axis (left) or fast/slow (right) jets. All sources are classified dichotomously as obscured/unobscured based on evidence for strong local X-ray absorption). Numbers in each quadrant indicate total population under each classification, symbols illustrate that population and accretor type.}\label{fig:precession_obscuration}
\end{figure*}

\section{Analysis}

In the previous section we have discussed obscured XRBs with reported evidence for strong and variable local absorption, which can dramatically affect the observed accretion luminosity. For none of the other sources in the FM25 sample is there reported evidence for such local absorption in excess with respect to the interstellar absorption (which is why for most multi-epoch X-ray analyses a single, fixed N$_H$, from the highest S/N fit, is generally adopted). In Table \ref{tab:data} we take the FM25 sample, but retaining just the columns relating to jet speed and whether or not there is evidence for jet precession. We then add a further column, which indicates whether or not there is evidence for strong local obscuration in the systems. Two small further changes have been made:  

\begin{itemize}
\item GRS 1915+105 now has two entries, one for its pre-obscured phase, i.e. before 2019, and the other for its obscured phase since. The former, fast/unobscured state corresponds to the entry in FM25; 
\item MAXI J1348-630 has an unconstrained speed, like several of the other BH XRBs. This is because Hughes et al. (2026, in prep) have recently demonstrated that deceleration model fits do {\em not} constrain the Lorentz factor for MAXI J1348-630.
\end{itemize}

For each system in our sample we are now able to classify it dichotomously as obscured/unobscured, precessing/fixed and fast/slow. These binary classifications of systems are illustrated in Fig. 1.  Monte Carlo sampling of the distances and inclinations from which speed limits are derived is not necessary in this context, as we demonstrated in FM25 that the resulting classifications and conclusions are robust to such uncertainties. In the following analysis, each entry in Table 1 is treated as an independent observational instance; in a small number of cases this corresponds to distinct accretion states of the same source.

\subsection{Statistical analysis}
\label{sec:stats}

To test for associations between jet properties and local obscuration
we constructed $2\times2$ contingency tables and analysed them using
two complementary approaches, described below.
The contingency tables -- testing association between obscuration and
precession, and obscuration and jet speed -- are illustrated in
Fig. 1.
Sources with completely uncertain classifications (i.e. "?" in Table 1) were excluded from the relevant
test. For sources with tentative classifications (e.g. ‘L?’), we adopt the indicated classification for the statistical analysis.

\subsubsection{Bayesian posterior probabilities}
We model the obscured fraction in each group as an independent
Bernoulli process \citep{Wall2012}.
Let $\theta_A$ and $\theta_B$ denote the true obscured fractions in
groups A and B respectively.
We place a prior distribution on each fraction using the Beta
distribution, $\mathrm{Beta}(\alpha, \beta)$, which is defined on
$[0,1]$ and has the form $p(\theta) \propto
\theta^{\alpha-1}(1-\theta)^{\beta-1}$; it is the natural prior
for a probability parameter and encompasses a wide range of beliefs,
from strongly informative to completely uniform.
Adopting uninformative Beta$(1,1)$ (uniform) priors, the posteriors
follow analytically from Beta--Binomial conjugacy:
\begin{equation}
  \theta_A \mid \mathrm{data} \sim \mathrm{Beta}(a+1,\,b+1), \quad
  \theta_B \mid \mathrm{data} \sim \mathrm{Beta}(c+1,\,d+1),
  \label{eq:posteriors}
\end{equation}
where $a$, $b$ are the counts of obscured and unobscured sources in
group A, and $c$, $d$ likewise for group B.
Crucially, given the choice of a Beta prior, these posteriors are
\emph{exactly} determined by the sample counts alone, with no
sampling or approximation required: the prior pseudo-counts simply
add to the observed counts to give the posterior parameters.
The posteriors are illustrated in Fig.~2 and listed in the figure
legend.

We evaluate the posterior probability
\begin{equation}
  P(\theta_A > \theta_B \mid \mathrm{data})
  = \int_0^1 f_A(x)\,F_B(x)\,\mathrm{d}x,
  \label{eq:prob}
\end{equation}
where $f_A$ is the Beta$(a+1, b+1)$ probability density function and
$F_B$ is the Beta$(c+1, d+1)$ cumulative distribution function.
This integral is evaluated by adaptive Gauss--Kronrod quadrature
\citep[as implemented in \textsc{scipy};][]{Virtanen2020}, which is
exact to numerical precision for the smooth, bounded integrands
arising here.
This is the direct probability that group A has a genuinely higher
obscured fraction than group B -- the quantity of scientific interest
-- expressed without recourse to Monte Carlo sampling.

For the direction test (precessing vs fixed-axis jets, $N=13$), we
find $P(\theta_{\rm prec} > \theta_{\rm fixed}) = 99.1$\%: strong
evidence that precessing jets are preferentially found in obscured
systems.
For the speed test (slow vs fast jets, $N=23$), we find
$P(\theta_{\rm slow} > \theta_{\rm fast}) = 97.3$\%: a weaker but
still suggestive trend in the same direction.

We note that the choice of prior has negligible influence on these
results.
Replacing the uniform Beta$(1,1)$ prior with the Jeffreys prior
Beta$(0.5, 0.5)$ -- which is the formally non-informative prior
invariant under reparametrisation of $\theta$ -- shifts the posterior
parameters to Beta$(a+0.5, b+0.5)$ for each group, since the Beta
family is conjugate for any Beta prior.
For our sample counts, this changes $P(\theta_A > \theta_B)$ by less
than $0.5$\% in both tests.

\subsubsection{Fisher's exact test}

As a cross-check we also applied Fisher's exact test \citep{Wall2012}
to both contingency tables.
This frequentist approach conditions on the marginal totals and
computes the probability of obtaining a table at least as extreme as
that observed under the null hypothesis of no association.
It is the appropriate frequentist test for small samples where the
$\chi^2$ approximation is unreliable.
The direction test yields OR $= 30.0$, $p = 0.029$ (significant at
the 5\% level), and the speed test yields OR $= 9.2$, $p = 0.069$
(marginally significant; ${\sim}1.8\sigma$ equivalent), where the
odds ratio OR quantifies how much more likely obscuration is in one
group compared to the other.
These results are consistent with the Bayesian analysis: with a
uniform prior and small samples the Bayesian posterior probability
and the frequentist $p$-value carry similar information, though the
former is more directly interpretable as a statement about the
hypothesis of interest.

\begin{figure*}
	\includegraphics[width=1.9\columnwidth]{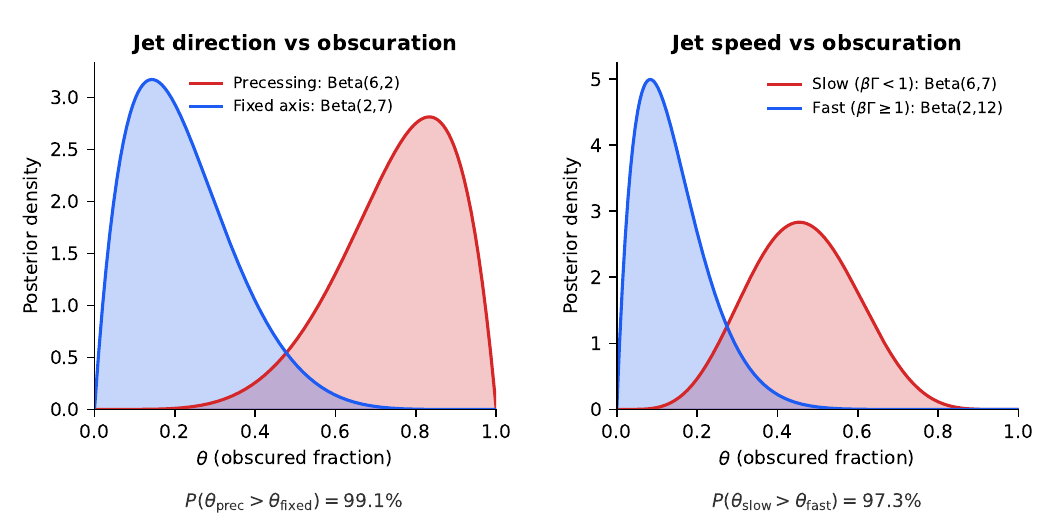}
\caption{Bayesian posterior distributions for the obscured fraction $\theta$ for (left) precessing (red) and fixed-axis (blue) jets; (right) slow ($\beta \Gamma < 1$, red) and fast ($\beta \Gamma \geq 1$, blue) jets. The significance of the difference between the posteriors in each case is given below the plots.}\label{fig:posteriors}
\end{figure*}

\section{Discussion}\label{sec:discussion}

In the following we use 'system' as shorthand for 'accretion system', i.e. we consider that for the few sources with more than one entry (e.g. GRS 1915+105 pre- and post-2019) those can be considered as independent when considering the connection between jet properties and obscuration.

Our results show that the association between precessing jets and obscured systems is statistically robust (Figs 1 \& 2, left panels). We emphasize that we only include in this analysis systems for which we are confident either that the jet axis is fixed or varying. There are two exceptions to the association, which we briefly address below. 

\begin{itemize}
    \item Sco X-1 appears to have precessing jets, but is not obscured. Sco X-1 contains a NS, so following FM25 (see also discussion below) it is unlikely to have fast jets. Sco X-1 is also known to accrete consistently close to the Eddington luminosity, and thus may not be too far away from the super-Eddington accretion inferred for some, maybe all, of the obscured systems. We note that \cite{Bradshaw2003} argued for branch-dependent absorption linked to different accretion regimes, and thus to changing line of sight into the disk/corona geometry.
    It is also worth stressing that Sco X-1 is a member of the 'Z sources' classification of NS XRBs\footnote{Neutron stars persistently accreting at (near-)Eddington rates, showing distinctive patterns of X-ray spectral and temporal variability, and strong and variable radio emission.}, an observational characteristic also shared by Cir X-1, which is both intermittently obscured and has slow, precessing, jets. It will be particularly interesting to determine whether other Z sources also exhibit jet precession, especially given the recent reports of dense winds in absorption from the Z source GX 13+1 \citep{Xrism2025}. Although no resolved jet speed exists for GX 13+1, this result strengthens the general picture that near-Eddington NS XRBs can produce dense obscuring winds. In this regard, Cyg X-2 represents perhaps the most promising candidate, as it is the only other Z source for which a resolved jet has been reported \citep{Spencer2013}.
    \item Cyg X-3 may represent a case in which occasional fast, fixed-axis jets are produced within an otherwise obscured system. However, the source undergoes significant secular variability, potentially driven by changes in the wind properties of the Wolf-Rayet donor star. It is therefore plausible that the infrequent fast ejections occur during episodes of reduced local obscuration. Future, strictly simultaneous radio and X-ray observations may clarify this issue.
\end{itemize}

\smallskip

The association between jet speed and obscuration is also very strong (Figs. 1 \& 2, right panels), and is based on a significantly larger sample of systems, as systems with only a single observed ejection can also be included in this case. It is striking that with the exception of the fast jets from Cyg X-3, {\em all} jets from obscured systems are slow, and {\em all} are precessing.

What is the origin of the association between obscured sources and slow jets? A plausible and straightforward explanation for the lower jet speeds is that jets in dense environments become mass-loaded very early, and hence are rapidly decelerated (see below for references to AGN works on this).
This seems especially plausible for SS433, where atomic emission lines are famously observed coming from the precessing jets (\citealt{Margon1984a}, \citealt{Blundell2011}). 
However, if this is the explanation, than a major open question remains: why have we only seen atomic emission lines from SS433, and not from other systems? Perhaps the slow, precessing, more or less persistent, jets in Cir X-1 \citep{Cowie2025} are the best candidate to test this, but the much longer inferred precession period will require a long campaign to reveal the moving lines. 

The deceleration of black hole jets due to mass loading is something which has previously been explored in the context of AGN and simulations (\citealt{Perucho2014b, Chatterjee2019}), and 
we note that recent simulations by Kwan, Dai et al. (in prep). show that in super-Eddington accretion scenarios, the higher the accretion rate, the slower the jets. Furthermore,  blast-wave modelling of the large-scale jets from the unobscured BH XRBs MAXI J1820+070, MAXI J1535-571, and XTE J1752-223 places upper limits on the local ISM density as low as $10^{-4}$ cm$^{-3}$ (\citealt{Carotenuto2024}; see also \citealt{Savard2025}, \citealt{Cooper2025}), directly supporting the picture that fast jets from unobscured X-ray binaries propagate into low-density environments. Despite this, the simulations of fast jets from unobscured sources such as MAXI J1820+070 do illustrate that entrainment of, and transfer of kinetic energy to, the ambient medium plays a role in their deceleration at larger scales \citep{Savard2025}. 

What, then, is the origin of jet precession in this scenario? In FM25 we noted that two scenarios are possible, one in which slower, precessing jets are launched from further out in the accretion flow than the fastest jets. In that scenario the launch region for the slow jets may itself be precessing, whereas the fastest jets arise from a region where the Bardeen-Petterson effect has aligned the accretion flow with the black hole spin. An alternative origin for precessing jets is that presented by \cite{Begelman2006} for SS433, in which the jet direction is determined by a massive precessing outflow that is able to successfully divert the jet. As we suggested in FM25, it seems plausible that for a number of the systems this scenario may be a viable alternative to the jets being launched from a (large) precessing region of the accretion flow. 
While difficult to demonstrate conclusively, it is plausible that most -- if not all -- of the obscured systems discussed above represent Galactic examples of super-Eddington accretion. This interpretation appears unambiguous in the case of SS433, and is strongly supported for the remaining systems. If this is the case, it is not surprising for outflowing discs to undergo precession. A super-Eddington disc would tend to be radiation-pressure dominated, which generally implies that it is also geometrically-thick. Such a disc can readily exhibit solid-body-like precession \citep[see, e.g.,][]{Ingram2009, Motta2018}. This has been essentially proven in the case of V404 Cyg, where a radiation-pressure, outflowing disk was also sustaining a rapidly (hours) precessing jet \citep{Miller-Jones2019}.

Interestingly, four of the six obscured sources are also recently-established VHE gamma-ray sources. Whether the super-Eddington accretion or the presence of a high density baryon-loaded environment is more important for a VHE detection remains to be seen.

It is important to note, as is clear in Figs. 1 \& 2,  that a wide range of {\em unobscured} sources, both NS and BH XRBs, also produce slow jets, and in fact the obscured fraction of systems with slow jets is close to 50\%. This could be consistent with a power-law distribution of black hole jet Lorentz factors \citep{Lilje2026} combined with a speed cap for resolved ejecta from NS XRBs. This implies that while a slow jet is not a marker of obscuration, in the presence of obscuration it is highly likely that the jets are slow (and precessing). 

An obvious implication of this analysis is that other obscured sources for which we do not yet have jet speed measurements should be producing slower, precessing, jets. While it is not clear that Her X-1 has a jet, and those in V4641 and GROJ1655-40 are associated with rare outburst states, GX 13+1 is a persistently variable and radio-bright system showing strong evidence for obscuration, which should be an excellent target for VLBI observations. Independent of high-resolution radio observations, measurement of mass-loading of relativistic ejecta may also be accessible via precision spectropolarimetic analysis of radio emission from XRBs (Hughes et al. in prep).

\subsection{Revisiting the Fender \& Motta 2025 result}

We can also apply precisely the same technique used here to the question of whether fast jets are preferentially locked to a fixed axis, as reported to be the case in FM25. Adopting the same approach as in section 3, we find that fast jets are more likely to be locked to a fixed axis than slow jets at 99.8\% confidence. The Monte Carlo approach in FM25 is more robust to uncertainties in distance and inclination, which affect the derived true jet speeds, but it is reassuring that the methods agree very closely. Adding the new measurement of a slower, precessing jet from GRS 1915+105 since 2019 \citep{Jiang2026} increases this to 99.9\%.

\section{Conclusions}

Starting from the sample considered in FM25 and adding in a classification of whether the accretion flow is obscured (absorbed) or not, and a small amount of additional data, we have shown empirically that there is a strong association between systems for which the accretion flow is heavily obscured, and slow, precessing jets. These systems are likely to be examples of (sometimes persistent, other times transient) super-Eddington accretion in our galaxy, and we speculate that this association may be due to early mass loading of the jets in unusually dense environments. 
Combining this with the result from FM25, a consistent picture emerges: the fastest black hole jets propagate along fixed axes in systems with low local absorption, whereas highly obscured systems preferentially produce slower, precessing jets.

\section*{Acknowledgements}

RF acknowledges support from the UKRI, European Research Council Synergy Grant 'Blackholistic' (grant agreement number 101071643) and The Hintze Family Charitable Foundation. RF thanks James Matthews and Fraser Cowie for useful discussions on this topic and a careful review of the manuscript.

\section*{Data Availability}

All the data used in the analysis in this paper are provided in Table 1 (see also Table 1 in Fender \& Motta 2025).



\bibliographystyle{mnras.bst}
\bibliography{biblio} 









\bsp	
\label{lastpage}
\end{document}